# Tuning Co valence state in cobalt oxyhydrate superconductor by post reduction

*Zhi Ren, Jian-lin Luo[¶], Zhu-an Xu, Guang-han Cao\**

Department of Physics, Zhejiang University, Hangzhou 310027, People's Republic of China

[¶]Beijing National Laboratory of Condensed Matter Physics, Institute of Physics, Chinese Academy of Sciences, Beijing 10080, People's Republic of China

*To whom correspondence should be addressed. E-mail: ghcao@zju.edu.cn. Tel/Fax: (86)571-87952590.

# ABSTRACT

We report a successful tuning of Co valence state in cobalt oxyhydrate superconductor via a facile post reduction using NaOH as reducing agent. The change in Co valence was precisely determined by measuring the volume of the released oxygen. The possible hydronium-incorporation was greatly suppressed in concentrated NaOH solution, making the absolute Co valence determinable. As a result, an updated superconducting phase diagram was obtained, which shows that the superconducting transition temperature increases monotonically with increasing Co valence in a narrow range from +3.58 to +3.65.

SYNOPSIS TOC

**Zhi Ren, Jian-lin Luo, Zhu-an Xu and Guang-han Cao\***

*Chem. Mater.* **200?**, *??*, ????

Tuning Co valence state in cobalt oxyhydrate superconductor by post reduction

The Co valence state ($V_{Co}$) in cobalt oxyhydrate superconductor was successfully tuned via post reduction using NaOH as reducing agent. The change in $V_{Co}$ can be determined precisely by measuring the volume of the released oxygen. Based on this approach, an updated superconducting phase diagram for the cobalt oxyhydrate superconductor has been established.

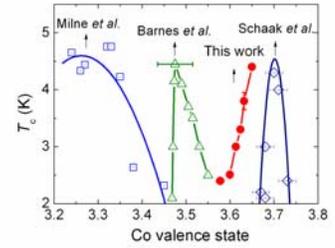

## Introduction

One of the major challenges in the newly discovered $Na_xCoO_2 \cdot yH_2O$ superconductor[1] is to establish experimentally the superconducting phase diagram (SPD), *i.e.*, the dependence of the superconducting critical temperature $T_c$ on the electron doping (or the Co valence state, $V_{Co}$). Schaak *et al.*[2] initially reported a narrow dome-shaped SPD with the optimal sodium content for the occurrence of superconductivity at $x$=0.3, corresponding to the formal Co valence of +3.7. However, the SPD based on sodium content alone is not well reproducible.[3] Using redox titration Milne *et al.*[4] proposed a revised SPD where the maximum $T_c$ was achieved at $V_{Co} \sim$ +3.3, which means that the superconductor is hole-doped rather than electron-doped. Very recently, Barnes *et al.*[5] suggested an asymmetric dome-like SPD where the optimal $V_{Co}$ was about +3.5. Besides, Takada *et al.*[6] showed that $T_c$ was also significantly affected by the isovalent exchange between $Na^+$ and $H_3O^+$, which further complicates the situation. In a word, the SPDs proposed by different groups are far from consistency.

The above controversial results basically come from the uncertainty in determining $V_{Co}$. First of all, the initial assumption that $V_{Co}$ is determined by sodium content alone is not strictly correct, because subsequent experimental results[7,8] indicated inevitable incorporation of hydronium ions during the hydration process. The possible existence of oxygen vacancies in $CoO_2$ layers also results in the "lower-than-expected" value of $V_{Co}$.[9] Secondly, wet-chemical redox analysis[3,9] is an alternative way to determine the Co oxidation state, however, complex side reactions and uncertainty of water content may introduce a large error for $V_{Co}$.[10] Thirdly, the incorporation of hydronium ions probably leads to phase separation (into Na-rich domains and Na-poor domains), which brings further uncertainty to establish a reliable SPD.

To minimize the measurement errors, we develop a new strategy to tune $V_{Co}$ utilizing soft-chemical reduction *after* the superconductor is synthesized. Based upon previous experimental results,[7,8] the post reduction in aqueous NaOH solution at room temperature can be expressed as follows,

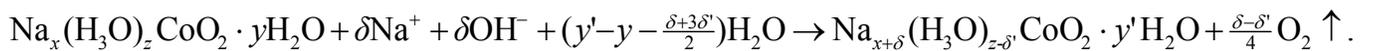

$$Na_x(H_3O)_zCoO_2 \cdot yH_2O + \delta Na^+ + \delta OH^- + (y'-y-\tfrac{\delta+3\delta'}{2})H_2O \rightarrow Na_{x+\delta}(H_3O)_{z-\delta'}CoO_2 \cdot y'H_2O + \tfrac{\delta-\delta'}{4}O_2 \uparrow.$$

Note that oxygen vacancy is ignored due to the following reasons: (1) both powder neutron diffraction[11] and recent redox titration analysis[12] indicate that the oxygen content of unhydrous $Na_{0.38}CoO_2$ is stoichiometric to a precision of 1 and 2 %; (2) even if there are some oxygen vacancies in $Na_{0.38}CoO_2$ they will be filled up by the oxygen from the intercalated $H_2O$[10]. From the above chemical equation, the change in $V_{Co}$ can be determined precisely by measuring the volume of the released oxygen, regardless of the extent of hydronium incorporations. It's emphasized that, in the presence of concentrated NaOH, the incorporated hydronium is mostly removed. Therefore, $V_{Co}$ of the heavily-reduced specimen can be determined merely by the sodium content, and the absolute $V_{Co}$ values of other successive samples can thus be given. By measuring the superconducting transition temperatures, as a result, an updated SPD for the cobalt oxyhydrate superconductor has been established.

**Experimental Section**

**Preparation of $Na_xCoO_2 \cdot yH_2O$**. Firstly, unhydrous compound $Na_xCoO_2$ was prepared by a solid state reaction method. High purity $Na_2CO_3$ (99.99%) and $Co_3O_4$ (99.99%) powders with the Na:Co ratio of 0.74:1 were mixed thoroughly and then pressed into pellets. The pellets were rapidly heated to 1073 K under flowing oxygen, holding for 12 hours. This process was repeated twice with intermediate grindings. Consequently pure $\gamma$-$Na_xCoO_2$ was obtained, as checked by powder x-ray diffractions (XRD). Secondly, partial sodium was deintercalated from the $\gamma$-$Na_xCoO_2$ phase using $Br_2/CH_3CN$ as oxidizing agent. To ensure a deep oxidation, the amount of bromine used was 30 times excess. The topotactic reaction was carried out in a sealed container at 313 K for five days. The resultant black powder was filtered, washed with acetone for several times and then dried in vacuum. Finally, the powder was placed in a chamber with the relative humidity of 100% at 300 K for two weeks for the thorough hydration. The resultant (hereafter called as-prepared superconductor) was preserved below 263 K in a refrigerator for the succeeding experiments.

**Post Reduction in Aqueous NaOH solutions.** Post reduction was carried out using apparatus shown in Figure 1. Before the experiment, NaOH solution was saturated with oxygen by flowing oxygen gas for 24 hours. The buret and the flask were washed with oxygen gas before the oxygen-saturated NaOH solution was infused into the flask and the buret. The as-prepared superconductor powder (about 0.5 g) was carefully wrapped with plastic film and then placed on the top the modified buret. After the sample had been immersed, the valve of the buret was shut down immediately. The reading of the liquid level of the buret was recorded as $V_1$ (ml). Once the sample contacts with the NaOH solution, small bubbles form on the sample's surface, and they were released to the top of the buret. The post reduction was carried out for enough time (24 ~ 48 hours) to ensure a complete reaction. The final reading of the liquid level was recorded as $V_2$ (ml). After the post reduction, the powder sample was filtered, washed with deionized water for several times until the pH value of the filtered solution was below 9. The resultant powder was then preserved below 263 K in a refrigerator. According to the reaction equation, the change in Co oxidation state can be calculated directly from the volume of the released oxygen by the following equation,

$$\Delta V_{Co} = 4 \times \frac{M}{m} \times \frac{P \times (V_1 - V_2)}{RT},$$

where $m$ is the mass of the as-prepared superconductor, $M$ the molecular weight of the initial $Na_x(H_3O)_zCoO_2 \cdot yH_2O$, $P$ the oxygen pressure (equals to the ambient pressure because the two liquid surfaces were kept at the same level while reading $V_1$ and $V_2$), $T$ the ambient temperature and $R$ the gas constant. $M$ can be calculated after $x$, $y$ and $z$ were determined by the inductively coupled plasma atomic emission spectroscope (ICP-AES) technique and the thermogravimetric (TG) analysis.[13] The dominant error of $\Delta V_{Co}$ comes from the volume measurement which was at most 0.04 ml. Considered that the $M$ value was 125±4 (this large uncertainty mainly comes from the variation of hydronium incorporation and the absorbed free water), the precision of $\Delta V_{Co}$ was still better than 0.004.

**Sample's Characterization.** Each sample was investigated by powder x-ray diffraction using a D/Max-rA Diffractometer with the Cu Kα radiations. Lattice parameters were refined by a least-squares fit with the consideration of the zero shifts. The Na:Co molar ratio of each sample was analyzed by the ICP-AES technique. The measurement precision was better than 2%. The thermogravimetric analysis was carried out on a WCT-2 differential thermal balance, operating at a heating rate of 10 K/min from 298 K to 1373 K. The temperature dependence of ac magnetization ($M_{ac}$) was measured on a Quantum Design PPMS facility. The applied ac magnetic field is $H_{ac}$=10 Oe, thus the ac susceptibility can be calculated by $\chi_{ac}= M_{ac} / H_{ac}$.

**Results and Discussion**

In our post-reduction experiments the concentration of NaOH was respectively set to be 0.1 M, 0.33M, 1 M, 3.3 M and 10 M, as listed in Table 1. After the reduction, each sample was examined by XRD and compositional measurements. Figure 2 shows the XRD pattern for the as-prepared superconductor as well as the post-treated specimens. As can be seen, the reduced samples show almost identical XRD pattern to that of the as-prepared superconductor, indicating that the post reduction maintains the crystal structure with bilayers of water [1]. At the same time, the sodium content in the post-reduced samples also increases with the concentration of NaOH (see Table 1), suggesting that partial $Na^+$ ions intercalate back into the layered structure. This result is consistent with the chemical equation proposed above. One can see in Table 1 that the water content surprisingly decreases with increasing Na content. A possible explanation is that, in the case of high Na content, one water molecule may serves as the ligand for two $Na^+$ ions. Detailed structural determination is called for this issue.

As shown in Table 1, the change of $V_{Co}$, $|\Delta V_{Co}|$, increases with increasing the concentration of NaOH, indicating a successful tuning of Co valence state. It is noted that $\Delta V_{Co}$ is not exactly equal to the value inferred from the sodium content alone. This is due to the variations of hydronium incorporation. In fact, $\Delta V_{Co}$ is determined by the changes of both sodium content and hydronium content, i.e., $\Delta V_{Co}=\delta'-\delta$. The hydronium content was measured by TG analysis, which shows that the sample treated in relatively concentrated NaOH ($c_{NaOH}\geq$1M) is almost free of hydronium (see Supporting Information). This is not surprising since $Na^+$ and $OH^-$ *cooperatively* suppress the hydronium incorporation. Therefore, the hydronium content is negligible in the presence of concentrated, e. g. 10 M, NaOH.

As the Na content of the sample treated in 10 M NaOH in Table 1 is 0.42, it is easy to see that the absolute value of $V_{Co}$ is +3.58. Accordingly, $V_{Co}$ values of the other samples were determined in the range from +3.6 to +3.65 using the $\Delta V_{Co}$ data. The result is in reasonable agreement with recent ARPES[14], NMR[15] and XAS[16] studies. With using obtained $V_{Co}$ data, as a matter of fact, the hydronium content of the samples can be calculated by the formula $z = 4 - x - V_{Co}$. The calculated $z$-value is in good agreement with the TG analysis within experimental errors.

Figure 3 shows the lattice parameters as a function of $\Delta V_{Co}$. On one hand, the $a$-axis reasonably expands slightly as $V_{Co}$ decreases. On the other hand, $c$-axis tends to shrink with the decrease of $V_{Co}$, in agreement with the strengthening of the interlayer Coulomb attraction between $Na^+$-ion layers and the negatively charged $CoO_2$ layers. However, there is one exception: the $c$-axis of the second sample surprisingly increases. We attribute this increase of $c$-axis to the increase of hydronium content as shown in Table 1. Previous studies[7,8,13] demonstrated that the $c$-axis became unusually large when the hydronium content was comparable to the sodium content. On the contrary, we found that the $c$-axes were almost identical and particularly short for the last three samples in Table 1. This fact further supports that the hydronium content becomes nearly zero under the circumstance of concentrated NaOH.

Figure 4 shows temperature dependence of ac magnetic susceptibility ($\chi_{ac}$) for each powder sample. The as-prepared superconductor does show bulk superconductivity at 4.5 K. The $\chi_{ac}$ value at 2 K achieves $-8 \times 10^{-3}$ emu/g, corresponding to a volume fraction of magnetic shielding as high as ~ 30% (this is among the best results ever reported for powder samples). The post-reduced samples also show superconducting transitions with clearly different $T_c$ ranging from 2.4 K to 3.8 K. One may note that the superconducting transition is much broader for the second sample, suggesting a high degree of inhomogeneity. As discussed above, the sample noticeably contains hydronium ions, which possibly leads to a kind of phase separation. Nevertheless, other samples show sharp superconducting transition, as can be seen in the inset of Figure 4.

Barnes *et al.*[5] recently reported a time-dependent superconducting properties for the cobalt oxyhydrate superconductor which was, however, not observed in the present study. The difference may arise from the different way in synthesizing the superconductor. They employed aqueous $Br_2$ solution to deintercalate the sodium from the parent compound $\gamma$-$Na_xCoO_2$. Since the aqueous $Br_2$ solution is acidic, the hydronium content of the synthesized cobalt oxyhydrate superconductor would be considerably high[13]. So, the reported "time-dependent superconductivity" seems to be due to the phase separation induced from the remarkable incorporation of hydronium. Besides, our samples were kept below 263 K, which slowed down the possible redox reaction that could alter $V_{Co}$ in the cobalt oxyhydrate superconductors.

Figure 5 shows our updated SPD plotted as a function of Co valence state. The superconducting region in our SPD is between those of Schaak *et al.*[2] and Barnes *et al.*[5] As can be seen, superconductivity appears in the vicinity of the quarter filling which show charge ordering in the unhydrous $Na_xCoO_2$ system[17]. The striking feature of our phase diagram is that $T_c$ decreases monotonously with decreasing $V_{Co}$. Actually, based on the neutron diffraction study Lynn *et al.*[18] has also suggested that $T_c$ increased as electrons transferred off the cobalt in the cobalt oxyhydrate superconductor, which is in agreement with our result. However, it is stressed that this is not a definitive conclusion for the shape of the SPD since we are unable to increase the $V_{Co}$ from the as-prepared superconductor. Moreover, synthesis of the "overdoped" superconductors is very difficult because such superconductor could be reduced by itself (the intercalated water) at room temperature.

According to the analysis by Shaak *et al.*,[2] $Na_xCoO_2 \cdot yH_2O$ would be a Mott-Hubbard insulator and a band insulator for $x=0$ ($Co^{4+}$) and $x=1$ ($Co^{3+}$), respectively, concerning the splitting of the Co 3d band due to the crystal field effect. Assuming a simple rigid band, in this scenario, each added Na above $x=0$ in $Na_xCoO_2 \cdot yH_2O$ results in the addition of one electron per cobalt to the upper Hubbard band. For $V_{Co} > 3.5$ as in the present result, the system will be electron-doped. One expects some careful experiments such as Hall measurement to make a definite conclusion on this issue.

**Conclusion**

In summary, we have presented a facile route to tune and measure the Co valence state in the cobalt oxyhydrate superconductor by a topotactic post reduction using NaOH as reducing agent. The remarkable advantage of this method lies in that the disturbance of hydronium incorporation can be eliminated to a great extent for the sample treated in concentrated NaOH, which makes the absolute Co valence determinable. Magnetic susceptibility measurement indicates that the superconducting transition temperature decreases monotonically with Co valence state in the range from +3.58 to +3.65. Further study is needed to clarify the doping dependence of $T_c$ for Co valence larger than +3.65. We expect that this approach of valence-tuning can also be applied to other related systems in the future.

ACKNOWLEDGMENT

We acknowledge the supports from National Science Foundation of China (No. 10674119 and No. 10674166) and National Basic Research Program of China (No. 2006CB601003).

**Supporting Information Available**: Thermal analysis of the specimens treated in 0.1 M and 1M NaOH. This material is available free of charge via the Internet at http://pubs.acs.org.

SUPPORTING INFORMATION PARAGRAPH

The thermal analysis for samples treated in 0.1 M and 1 M NaOH is shown in figure S 1. According to our previous analysis,[1] a small weight-loss around 481 K in figure S 1(a) accompanied by an endothermal peak is due to the loss of water from the incorporated hydronium ions. Therefore, the hydronium content is then determined to be 0.04(1) by the TG data. In figure S 1(b) for the sample treated in 1 M NaOH, in contrast, such a weight loss cannot be detected, which indicates that the hydronium content for the sample is, even if exists, negligible.

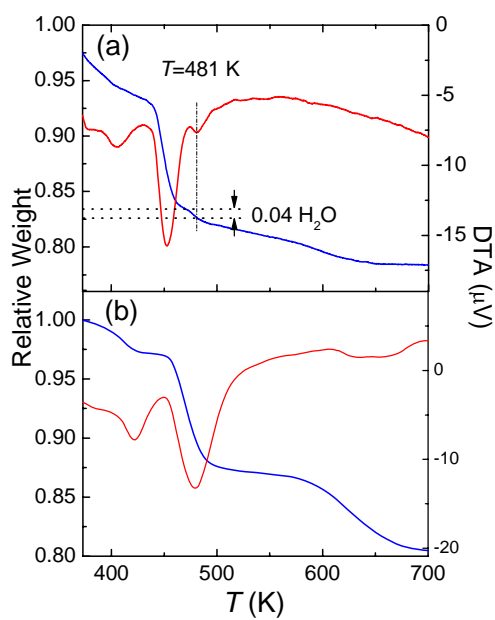

Figure S 1. Thermal analysis of the cobalt oxyhydrate superconductor post-reduced in (a) 0.1 M NaOH (b) 1 M NaOH.

(1) Ren, Z.; Wang, Y. W.; Liu, S.; Wang, J.; Xu, Z. A.; Cao, G. H. *Chem. Mater*. **2005**, *17*, 1501.

FIGURE CAPTIONS:

**Figure 1.** Experimental apparatus for measuring the amount of the released oxygen in the post reduction for the as-prepared cobalt oxyhydrate superconductor.

**Figure 2.** Powder x-ray diffraction for the as-prepared cobalt oxyhydrate superconductor as well as those post-treated in aqueous NaOH solutions. The inset shows an enlarged view of the (006) reflections.

**Figure 3.** The lattice parameters plotted as a function of change in Co valence ($\Delta V_{Co}$).

**Figure 4.** Temperature dependence of ac magnetic susceptibility for the cobalt oxyhydrate superconductors. As labeled, different concentration of NaOH solution was used to reduce the Co valence. The inset shows an enlargement of the superconducting transitions.

**Figure 5.** An updated superconducting phase diagram for cobalt oxyhydrate superconductors. For comparison, previous results proposed by some other groups are also presented.

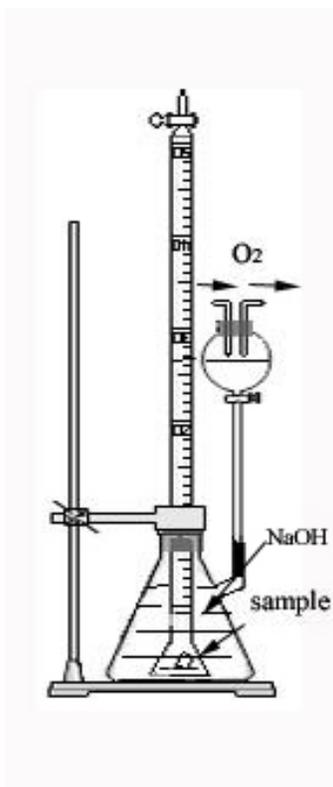

Figure 1. Z. Ren et al., submitted to *Chemistry of Materials*.

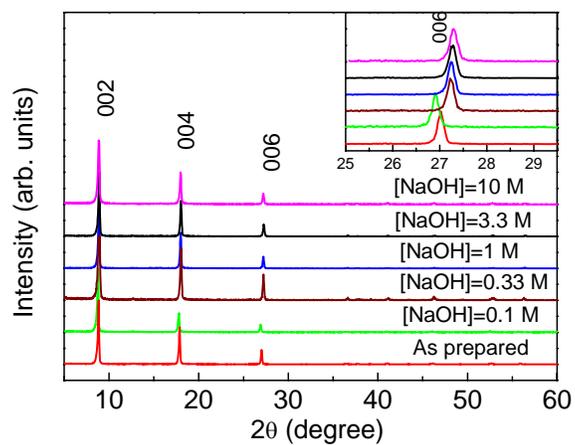

Figure 2. Z. Ren et al., submitted to *Chemistry of Materials*.

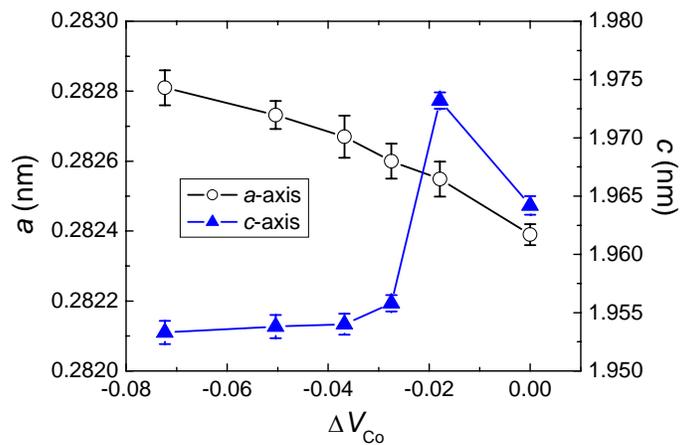

Figure 3. Z. Ren et al., submitted to *Chemistry of Materials*.

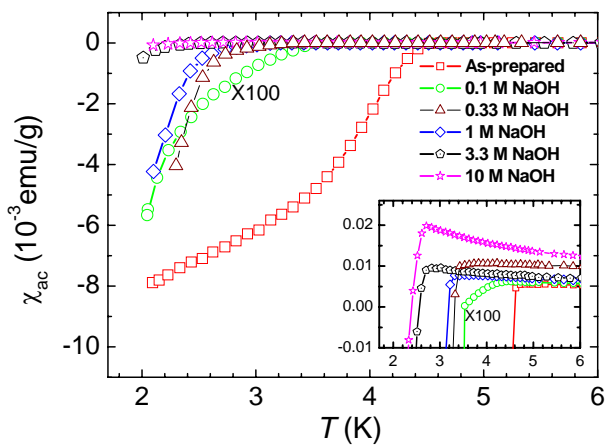

Figure 4. Z. Ren et al., submitted to *Chemistry of Materials*.

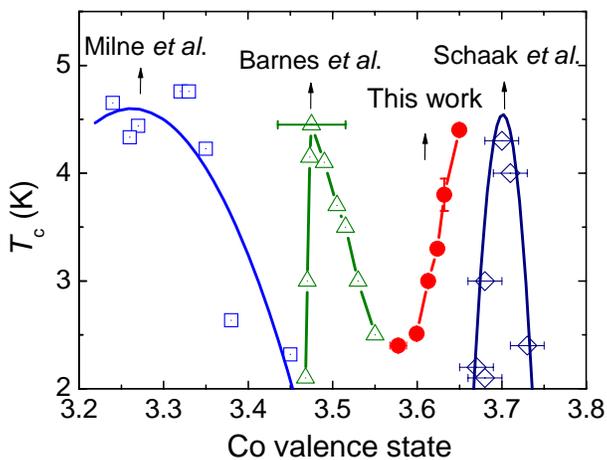

Figure 5. Z. Ren et al., submitted to *Chemistry of Materials*.

**Table 1**. Characterizations of the as-prepared $Na_x(H_3O)_zCoO_2 \cdot yH_2O$ superconductor as well as those post-treated in NaOH solutions.

| $C_{NaOH}$ | $x$ | $y$ | $z$ | $\Delta V_{Co}$ | $a$ (Å) | $c$ (Å) | $T_c$ (K) |
|---|---|---|---|---|---|---|---|
| 0 | 0.325 | 1.44 | 0.02 | 0 | 2.8239 | 19.642 | 4.5 |
| 0.1 | 0.33 | 1.19 | 0.04 | -0.018(2) | 2.8255 | 19.732 | 3.8 |
| 0.33 | 0.38 | 1.00 | 0.00 | -0.028(3) | 2.8260 | 19.558 | 3.2 |
| 1 | 0.387 | 0.93 | 0.00 | -0.037(3) | 2.8267 | 19.540 | 3.0 |
| 3.3 | 0.40 | 0.89 | 0.00 | -0.050(3) | 2.8273 | 19.538 | 2.5 |
| 10 | 0.42 | 0.90 | 0.00 | -0.072(4) | 2.8281 | 19.533 | 2.4 |